\documentclass[final,5p,times,twocolumn]{elsarticle}

\usepackage{graphicx}

\usepackage{amssymb}

\journal{Physica E}
\begin{document}

\begin{frontmatter}



\title{Collapse of the fractional quantum Hall state by 
a unidirectional periodic potential modulation}


\author[ISSP]{A. Endo}
\author[Tohoku]{N. Shibata}
\author[ISSP]{Y. Iye}

\address[ISSP]{Institute for Solid State Physics, University of Tokyo, Kashiwa, Chiba, 277-8581, Japan}
\address[Tohoku]{Department of Physics, Tohoku University, Sendai, Miyagi, 980-8578, Japan}

\begin{abstract}
We have investigated the effect of unidirectional periodic potential modulation on the fractional quantum Hall (FQH) states located within the filling factor range $1<\nu<2$. We find that odd-numerator FQH states are strongly suppressed, while the even-numerator ones are less affected or even enhanced, by the introduction of the modulation. In the picture of the composite fermions (CFs), the behaviors are equivalent to the suppression of the spin splitting of the CFs by the modulation. We discuss the origin of the suppression and the possible modulation-induced phase transition from the FQH state to the stripe state. 
\end{abstract}


\end{frontmatter}


The quantum Hall (QH) state is an incompressible state with a gap  in the energy spectrum. The gap $\Delta$ derives from the Landau quantization or the Zeeman splitting in the integer quantum Hall (IQH) state, while $\Delta$ opens up due to the strong electron interactions in the fractional quantum Hall (FQH) states (see, e.g., \cite{ChakrabortyPietilainen95}). In either of IQH and FQH states, $\Delta$ is envisaged to take the maximal value in an ideal two-dimensional electron gas (2DEG) devoid of disorders. In a real 2DEG, disorders are inevitably present and induce random fluctuations in the potential profile, resulting in a reduced gap, thereby making the QH states less robust. The values of $\Delta$ in an actual 2DEG are thus crucially affected by the magnitude and the spatial extent of the random potential. Unfortunately, however, it is usually difficult to know the profile of the random potential in detail, especially in a GaAs/AlGaAs system containing the 2DEG plane buried at the depth of typically 100 nm. It is therefore beneficial to study the behavior of $\Delta$ in a well-known potential profile. Unidirectional periodic potential modulation artificially introduced to a 2DEG is well-suited for this purpose; the commensurability oscillation (CO) \cite{Weiss89} observed in low magnetic fields allows us to determine the modulation amplitude with high accuracy \cite{Endo00e}. Of course, it should be born in mind that the spatial and orientational order in the unidirectional periodic modulation can have effects that are absent in the random potential.    
Using a unidirectional lateral superlattice (LSL) sample, we have studied the effect of the modulation on the IQH states \cite{Endo09IQHEMOD}. We found that $\Delta$ is reduced by the width acquired by the Landau levels (LLs) due to the modulation; the width oscillates with the magnetic field owing to the commensurability effect.
In the present paper, we focus on the FQH states observed in the filling factor range $1<\nu<2$. We will show that the effect of the modulation is quite different between odd-numerator and even-numerator states. 

\begin{figure}[tb]
\includegraphics[width=6.5cm]{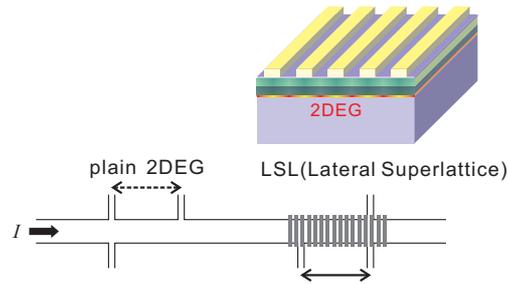}%
\caption{Schematic drawing of the Hall bar sample containing the LSL and the unpatterned plain 2DEG sections in series, with a sketch of the LSL depicted on the top right.\label{sample}}
\end{figure}

We used a conventional GaAs/AlGaAs 2DEG wafer with the heterointerface residing at the depth 90 nm from the surface. The electron density $n_e$ and the mobility $\mu$ were 2.9$\times$10$^{15}$ m$^{-2}$ and 106 m$^2$/(Vs), respectively, obtained after LED illumination. The potential modulation with a period $a$=184 nm was introduced by exploiting the strain-induced piezoelectric effect \cite{Skuras97} generated, on cooling, by a grating of the negative electron-beam resist placed on the surface (see Fig.\ \ref{sample}) \cite{Endo00e,Endo05HH}. We employed a Hall bar containing sections with (LSL) and without (plain 2DEG) the modulation in series, as depicted in Fig.\ \ref{sample}, which allowed us to discern the effects brought about by the modulation by comparing the two sections. By the analysis of CO, we found the amplitude of the modulation $V(x)=V_0\cos(2\pi x/a)$ to be $V_0$=0.31 meV\@. Measurements were carried out in a dilution refrigerator, with standard low-frequency ac lock-in technique (0.5 nA, 13 Hz).

\begin{figure}[tb]
\includegraphics[bbllx=30,bblly=90,bburx=500,bbury=765,width=8.0cm]{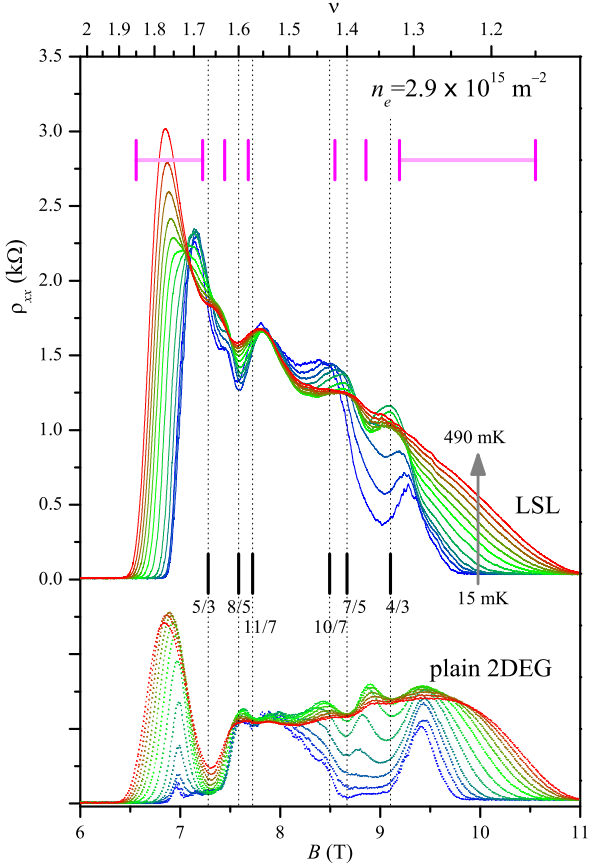}%
\caption{Longitudinal resistivity $\rho_{xx}$ at temperatures ranging from 15 mK (bottom, blue) to 490 mK (top, red) for the LSL (solid lines) and the adjacent plain 2DEG (dotted lines). Locations of the FQH states are indicated by vertical dotted lines. Vertical (magenta) ticks on the top denote the filling factors at which stripe correlation is predicted \cite{Shibata03} to be enhanced.\label{rhoxx}}
\end{figure}

\begin{figure}[tb]
\includegraphics[bbllx=30,bblly=50,bburx=560,bbury=770,width=8.5cm]{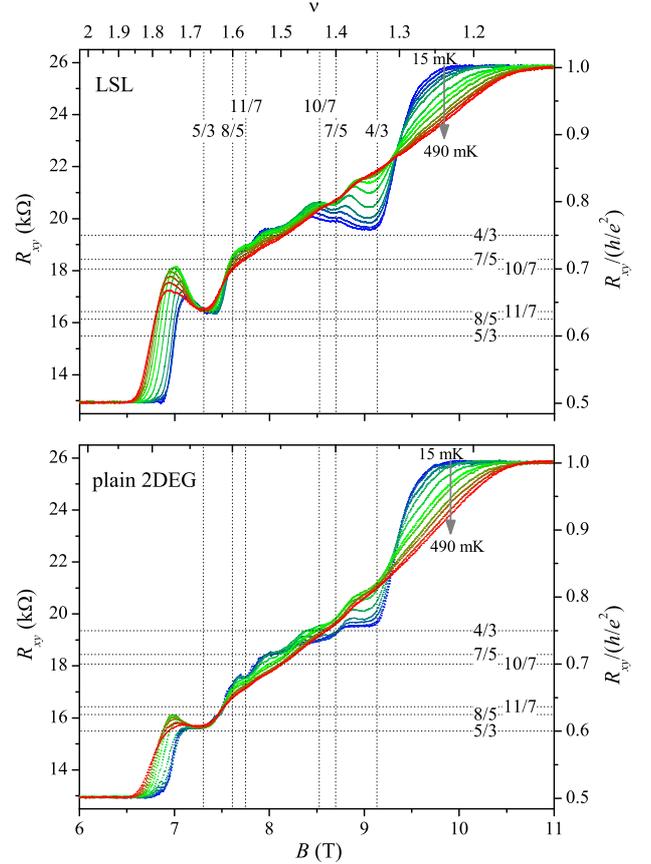}%
\caption{Hall resistance $R_{xy}$ at temperatures ranging from 15 mK (blue) to 490 mK (red) for the LSL (solid lines) and the adjacent plain 2DEG (dotted lines). Expected locations and the values of the FQH plateaus are indicated by vertical and horizontal dotted lines, respectively.\label{rhoxy}}
\end{figure}

In Figs.\ \ref{rhoxx} and \ref{rhoxy}, we show the traces of the longitudinal resistivity $\rho_{xx}$ and the Hall resistance $R_{xy}$, respectively, measured at temperatures ranging from 15 to 490 mK\@. In both figures, the traces from the LSL (the plain 2DEG) are plotted by solid (dotted) lines in the top (bottom) panels. First, we focus on $\rho_{xx}$. The plain 2DEG exhibits clear FQH resistivity minima that deepen with decreasing temperature at filling factors $\nu$=$5/3$, $7/5$, and $4/3$, and also a less clear but discernible minimum at $\nu$=$11/7$. Introduction of the modulation drastically alters the appearance of these minima. The minimum at $\nu$=$5/3$ completely disappears and a peak that grows with decreasing temperature appears instead at a slightly lower magnetic field. The minima at $\nu$=$7/5$ and $11/7$ disappear as well. In stark contrast, the minimum at $\nu$=$4/3$ survives in the presence of the modulation. Furthermore a new minimum develops at $\nu$=$8/5$, which is absent in the plain 2DEG\@. In brief, FQH states with odd numerator are destroyed by the modulation, while ones with even numerator withstand (or even reinforced by, as in $\nu$=$8/5$) the disturbance due to the modulation.

\begin{figure}[tb]
\includegraphics[bbllx=10,bblly=20,bburx=810,bbury=530,width=8.5cm]{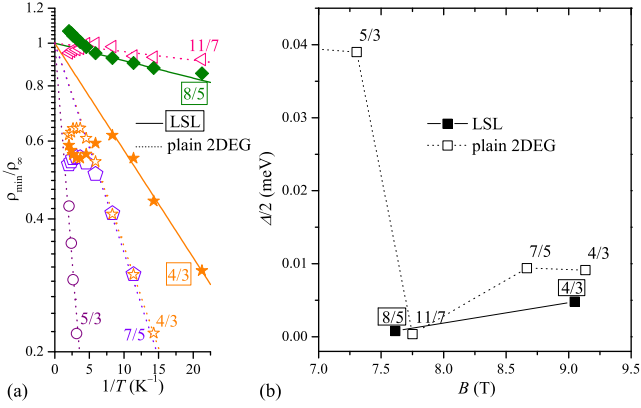}%
\caption{(a) Arrhenius plots for FQH states. Fits to Eq.\ (\ref{rhomin}) is also shown. (b) Activation energies deduced from the slopes of the linear fits shown in (a) plotted against the magnetic field. Solid and open symbols represent LSL and the plain 2DEG, respectively, in both figures.\label{actvarr}}
\end{figure}

To be more quantitative, we acquire the activation energies of the FQH states from their temperature dependence. As can be seen in the Arrhenius plots, Fig.\ \ref{actvarr} (a), the temperature dependence of the resistivity minima $\rho_\mathrm{min}$ at the FQH states are well described by
\begin{equation}
\rho_\mathrm{min} (T)= \rho_{\infty} \exp \left(-\frac{\Delta}{2 k_\mathrm{B} T} \right)
\label{rhomin}
\end{equation} 
for the range of the temperature shown in the figure. The $\Delta/2$ obtained from the fit are plotted as a function of the magnetic field in Fig.\ \ref{actvarr} (b). It is interesting to point out that, in the plain 2DEG, $\nu$=$4/3$ possesses smaller $\Delta/2$ than $\nu$=$5/3$, despite its better robustness against the modulation.

The effect of the modulation is less evident in the $R_{xy}$, Fig.\ \ref{rhoxy}. However, the plateau at $B\sim$7.4 T takes place in the LSL at higher value and at slightly higher magnetic field compared with that in the plain 2DEG, suggesting that the plateau derives from $\nu$=$8/5$ rather than from $\nu$=$5/3$. Again, the difference is less obvious for the $\nu$=$4/3$ plateau.

The difference we have found in the response to the modulation between odd- and even-numerator FQH states can naturally be related to the difference in the origin of the gaps. In the language of the composite fermions (CFs) \cite{JainBook07}, a FQH state at a filling factor $\nu$ in the range $1<\nu<2$ is interpreted as the IQH state at the filling factor $p=\pm (2-\nu)/(2\nu-3)$ of the CFs, where the $+$ sign ($-$ sign) applies for $\nu>3/2$ ($\nu<3/2$). Here, the CF is constructed by attaching two flux quanta to a hole (a deficit of electrons) from the $\nu$=2 state (the closed shell of the lowest Landau level (LLL) with both up- and down-spin branches fully occupied) \cite{Du95,Endo01c}. Thus, $\nu$=$5/3$ is mapped to $p$=1, $\nu$=$4/3$ and $8/5$ to $p$=2, and $\nu$=$7/5$ and $11/7$ to $p$=3; odd- and even-numerator FQH states correspond to odd- and even-integer QH states for the CFs, respectively. Therefore, by analogy to the IQH states of the electrons, the energy gaps for the odd-numerator and even-numerator FQH states observed in the present study are ascribable to $E_\mathrm{z}$ and $\hbar \omega_\mathrm{c}^\mathrm{CF}-E_\mathrm{z}$, respectively, in the magnetic field range studied in the present work, where $\hbar \omega_\mathrm{c}^\mathrm{CF} > E_\mathrm{z}$.
Here, $\hbar \omega_\mathrm{c}^\mathrm{CF}$ represents the cyclotron energy of the CFs, and $E_\mathrm{z} = g^{*} \mu_\mathrm{B} B$ is the Zeeman energy with the asterisk denoting the (possible) enhancement by the exchange effect.
The spin configuration implied in this assignment is in basic agreement with an early experiment \cite{Sachrajda90} carried out prior to the CF theory.
Note, however, that $E_\mathrm{z}$ can surpass $\hbar \omega_\mathrm{c}^\mathrm{CF}$ for large enough $B$ or $p$, since $\hbar \omega_\mathrm{c}^\mathrm{CF} \propto \sqrt{B}/(2p\pm 1)$ and $E_\mathrm{z} \propto B$ \cite{JainBook07}, or by introducing the in-plane component of the magnetic field by tilting, which enhances only $E_\mathrm{z}$ \cite{Clark89,Du95,Du97}. The changes we have perceived in the FQH states brought about by the modulation can thus be consistently understood by assuming that the modulation suppresses the Zeeman splitting of the CFs. In what follows, we discuss the possible mechanisms through which the Zeeman energy is reduced.

First, we discuss the effect of the strain. It is well known that the $g$-factor is reduced by the application of the hydrostatic pressure \cite{Morawicz93,Nicholas96}, namely with the decrease of the interatomic distances. In fact, decrease (increase) of the activation energies with increasing pressure was experimentally observed for odd-numerator (even numerator) FQH states by applying the hydrostatic pressure $\sim$5$-$20 kbar \cite{Morawicz93,Nicholas96}. As mentioned earlier, the modulation in our sample is introduced exploiting the strain generated by the electron-beam resist. The (compressive) strain induced by the resist should, in principle, have similar effect as the pressure. We can estimate \cite{Endo05HH}, with the aid of the theory by Larkin \textit{et al.}\cite{Larkin97}, the contraction (or a negative value of the dilation $\delta$) from the observed modulation amplitude $V_0$=0.31 meV\@. We find $\delta \sim -6\times 10^{-5}$ at the 2DEG plane (attenuated from $\delta \sim -2\times 10^{-3}$ at the surface). Using the bulk modulus of GaAs $\sim$80 GPa as the conversion factor, the values of $\delta$ are translated to the pressure of 0.04 kbar (1 kbar), which are much smaller than the pressure used in the experiment mentioned above; the reduction of the $g$-factor by the strain appears to be too small to explain the observed changes in the activation energies of the FQH states.

The second candidate is the suppression of the exchange-enhancement by the modulation. Values of the $g$-factors larger than bare $g$-factor in the bulk GaAs ($|g_\mathrm{GaAs}|$=$0.44$) are reported for the FQH states in $1<\nu<2$ by Du \textit{et al.} \cite{Du95,Du97}, implying that the Zeeman splitting is enhanced by the exchange effect. They showed that  $g^*$ varies roughly linearly with the effective magnetic field $B_\mathrm{eff}$ (see below), or as $g^*=g_0 [1+I (2-\nu)/\nu]$ with $g_0$=$0.42$ and $I$=1.5, and concluded that the $g$-factor of the CFs is largely deriving from the electron component of the particle. Note that the $g^*$ extrapolates to $g^*\sim g_0 \simeq |g_\mathrm{GaAs}|$ and 1.0 at $\nu=2$ and $\nu=1$, respectively, as expected for the electrons. The introduction of the modulation has been shown \cite{Manolescu95,Petit97} to be able to destroy the exchange enhancement by lifting the degeneracy of the LLs, which can be the mechanism for the suppression of the $g$-factor in our sample. The suppression, however, requires the energy dispersion within the extent of the wave functions to be large enough and comparable to the Coulomb energy, $\ell |dV/dx| \sim e^2/(4\pi \epsilon \epsilon_0 \ell)$. Here $\ell$=$\sqrt{\hbar/eB}$ is the magnetic length, which roughly represents the size of the wavefunctions in LLL\@. In our sample, $\ell |dV/dx| \sim$ 0.1 meV, and therefore the amplitude may be too small and the potential gradient may be too smooth for this mechanism to be fully operative.


We next turn our eyes to a density matrix renormalization group (DMRG) calculation by Shibata and Yoshioka \cite{Shibata03} on the ground state of the LLL\@. The theory reveals the enhancement of the stripe correlation, reminiscent of the stripe phase in the higher LLs \cite{Fogler96,Moessner96}, at around $\nu\sim$0.42, 0.37, and in the range 0.32$-$0.15, and at the fillings related to them by the particle-hole symmetry. Although the stripe phase in the LLL still lacks unambiguous experimental evidence, it seems quite possible that the stripe correlation is further reinforced by the introduction of a unidirectional modulation. In fact, in the second LL, we observed strongly anisotropic features at $\nu$=$5/2$ and $7/2$ in LSLs, which we attributed to the manifestation of the stripe phase stabilized by the modulation \cite{Endo02f}. The predicted positions for the enhancement of the stripe correlation are indicated in Fig.\ \ref{rhoxx}, with which the locations of the most of the peaks (or small humps) in the low-temperature traces of LSL are seen to coincide. This suggests a scenario that, in the LSL, the FQH states are destabilized by the modulation and superseded by the stripe states. The coincidence in the locations should be interpreted with care, however, since it may simply be resulting from the fact that they both take place interspersed with the FQH states. To further examine the  scenario, we performed preliminary DMRG calculations of the ground and excited states at $\nu$=$1/3$ in a 2DEG subjected to a periodic modulation, and compared with the behavior of $\nu$=$5/3$ (equivalent to $\nu$=$1/3$ by the particle-hole symmetry) in the experiment. The results will be presented elsewhere \cite{Endo09FQHEMOD}. Briefly, the calculation showed the phase transition from the FQH state to the stripe (unidirectional charge-density-wave) state having a period $\sim 4 \ell$, with the increase of the modulation amplitude. The $\nu$=$5/3$ FQH state was observed to appear or disappear in LSLs with differing modulation amplitudes in accordance with the calculation. The possible difference of this transition between odd-numerator and even-numerator states is yet to be studied. 

Finally, we comment on the CO of the CFs \cite{Smet99,Willett99,Endo01c}. It is well known that CFs behave as real particles following trajectories dictated by the effective magnetic field $B_\mathrm{eff}$, allowing for ranges of phenomena resulting from semiclassical cyclotron motion of the CFs \cite{JainBook07}. For CFs around $\nu$=$3/2$ ($B$=$B_{3/2}$), $B_\mathrm{eff}=-3(B-B_{3/2})$. We have actually observed CO of the CFs subjected to a unidirectional periodic potential modulation (which operate mainly as a modulation of the effective magnetic field) for CFs around $\nu$=$3/2$, using a LSL with a period $a$=92 nm \cite{Endo01c}. For such phenomena to be resolved, the mean free path of the CFs $L_\mathrm{CF}$ must be much larger than the artificially introduced length scale, the period $a$ in LSLs. In the present sample, $L_\mathrm{CF}=(h/e^2)(1/\rho_{3/2})\sqrt{3/(\pi n_e)}$=300 nm estimated from the resistivity $\rho_{3/2}$ at $\nu$=$3/2$, which is not so large compared with the period $a$=184 nm. Therefore CO is unlikely to be observed in the present sample. Furthermore, CO, if ever takes place, is limited to the proximity of $\nu$=$3/2$ in our sample with rather large $a$; the minima are expected to take place in the range 7.9 T $<B<$ 8.3 T\@. The range does not extend out into the magnetic field region where FQH states are observed. Therefore we conclude that the FQH states observed in the present study are unaffected by the commensurability effect of the CFs.

To summarize, unidirectional periodic potential modulation suppresses selectively the odd-numerator FQH states, consistent with the suppression of the CF spin splitting by the modulation. The collapse of the FQH states may possibly be related to the modulation-induced phase transition from the FQH state to the stripe phase, predicted by the DMRG calculation.

This work was supported by Grant-in-Aid for Scientific Research (C) (18540312) and (A) (18204029) from the Ministry of Education, Culture, Sports, Science and Technology (MEXT).








\bibliography{lsls,ourpps,qhe,ninehlvs,twodeg,cfs}

\end{document}